\documentclass[pra,showpacs,preprintnumbers,tightenlines,superscriptaddress,twocolumn]{revtex4}
\usepackage[pdftex]{graphicx}
\usepackage{dcolumn}
\usepackage{bm}
\usepackage{epsfig}
\usepackage{stmaryrd}
\usepackage{dsfont}
\usepackage{bbm}
\usepackage{amsmath}
\usepackage{amsfonts}
\usepackage{amssymb}
\usepackage{amscd}
\usepackage{amstext}
\usepackage{float}
\usepackage{rotating,standalone}
\usepackage{color}
\usepackage{ulem}
\usepackage[percent]{overpic}
\newcommand{\expec}[1]{\langle #1 \rangle}

\newcommand{\bra}[1]{\langle\,{#1}\, |}
\newcommand{\ket}[1]{|\,{#1}\,\rangle}
\newcommand{\braket}[2]{\mbox{$\langle\,{#1}\, | \,{#2}\,\rangle$}}

\newcommand{\tabvspace}{\Big.}

\newcommand{\id}{\mathds{1}}
\newcommand{\sub}[2]{{#1}_{\mbox{\!\! \scriptsize #2}}}

\newcommand{\bv}[1]{\mathbf{ #1 }}

\def\beq{\begin{equation}}
\def\eeq{\end{equation}}

\def\CR{\nonumber\\[0.15cm]}

\newcommand{\fref}[1]{FIG.~\ref{#1}}

\newcommand{\frefp}[2]{FIG.~\ref{#1}#2}

\newcommand{\eref}[1]{Eq.~(\ref{#1})}

\newcommand{\sref}[1]{section~\ref{#1}}

\newcommand{\cref}[1]{chapter~\ref{#1}}

\newcommand{\Cref}[1]{Chapter~\ref{#1}}
\newcommand{\tref}[1]{table~\ref{#1}}

\newcommand{\aref}[1]{appendix~\ref{#1}}
\newcommand{\bref}[1]{(\ref{#1})}

\newcommand{\ma}[1]{{\color{green} {#1} }}

\begin{document}
\title{Transport on flexible Rydberg aggregates using circular states}
\author{M.~M.~Aliyu}
\affiliation{Department of Physics, Bilkent University, Ankara 06800, Turkey}
\author{A.~Ulug{\"o}l}
\affiliation{Department of Physics, Bilkent University, Ankara 06800, Turkey}
\author{G.~Abumwis}
\affiliation{Department of Physics, Bilkent University, Ankara 06800, Turkey}
\affiliation{Max Planck Institute for the Physics of Complex Systems, N\"othnitzer Strasse 38, 01187 Dresden, Germany}
\author{S.~W\"uster}
\affiliation{Department of Physics, Bilkent University, Ankara 06800, Turkey}
\affiliation{Department of Physics, Indian Institute of Science Education and Research, Bhopal, Madhya Pradesh 462 023, India}
\email{sebastian@iiserb.ac.in}
\begin{abstract}
Assemblies of interacting Rydberg atoms show promise for the quantum simulation of transport phenomena, quantum chemistry and condensed matter systems.
Such schemes are typically limited by the finite lifetime of Rydberg states. Circular Rydberg states have the longest lifetimes among Rydberg states but lack the energetic isolation in the spectrum characteristic of low angular momentum states.
The latter is required to obtain simple transport models with few electronic states per atom. Simple models can however even be realized with circular states, by exploiting dipole-dipole selection rules or external fields.
We show here that this approach can be particularly fruitful for scenarios where quantum transport is coupled to atomic motion, such as adiabatic excitation transport or quantum simulations of electron-phonon coupling in light harvesting.
Additionally, we explore practical limitations of flexible Rydberg aggregates with circular states and to which extent interactions among circular Rydberg atoms can be described using classical models.
\end{abstract}

\maketitle

\section{Introduction}

We refer to flexible Rydberg aggregates as assemblies of Rydberg atoms that exhibit excitation transport or collective exciton states and are mobile in a possibly restricted geometry \cite{wuester:review}. They exhibit links between motion, excitation transport and coherence \cite{wuester:cradle,moebius:cradle,wuester:immcrad,schoenleber:immag} and spatially inflated Born-Oppenheimer surfaces for the simulation of characteristic phenomena from the nuclear dynamics of complex molecules \cite{wuester:CI,leonhardt:switch,leonhardt:orthogonal,leonhardt:unconstrained,zoubi:VdWagg}.

Most related experiments \cite{barredo:trimeragg,labuhn:rydberg:ising,cenap:emergent,maxwell:polaritonstorage,guenter:EITexpt,pineiro_orioli_spinrelax} and theory in this direction have so far focussed on aggregates based on Rydberg states with low angular momenta, $l=0,1,2$, due to the possibility of direct excitation and the energetic isolation provided by the energy gap to the nearest other states. For example~$|E(\ket{n=49,d}) -E(\ket{n=50,p})| =18.9 $ GHz in $^{87}$ Rb, which can be much larger than energy scales accessible by Rydberg aggregate dynamics. Here $n$ is the principal quantum number. However inertia and spontaneous decay limit realistic flexible Rydberg aggregate sizes to less than $\sim 4-10$ atoms for these low angular momentum states.%

\begin{figure}[htb]
\centering
\epsfig{file=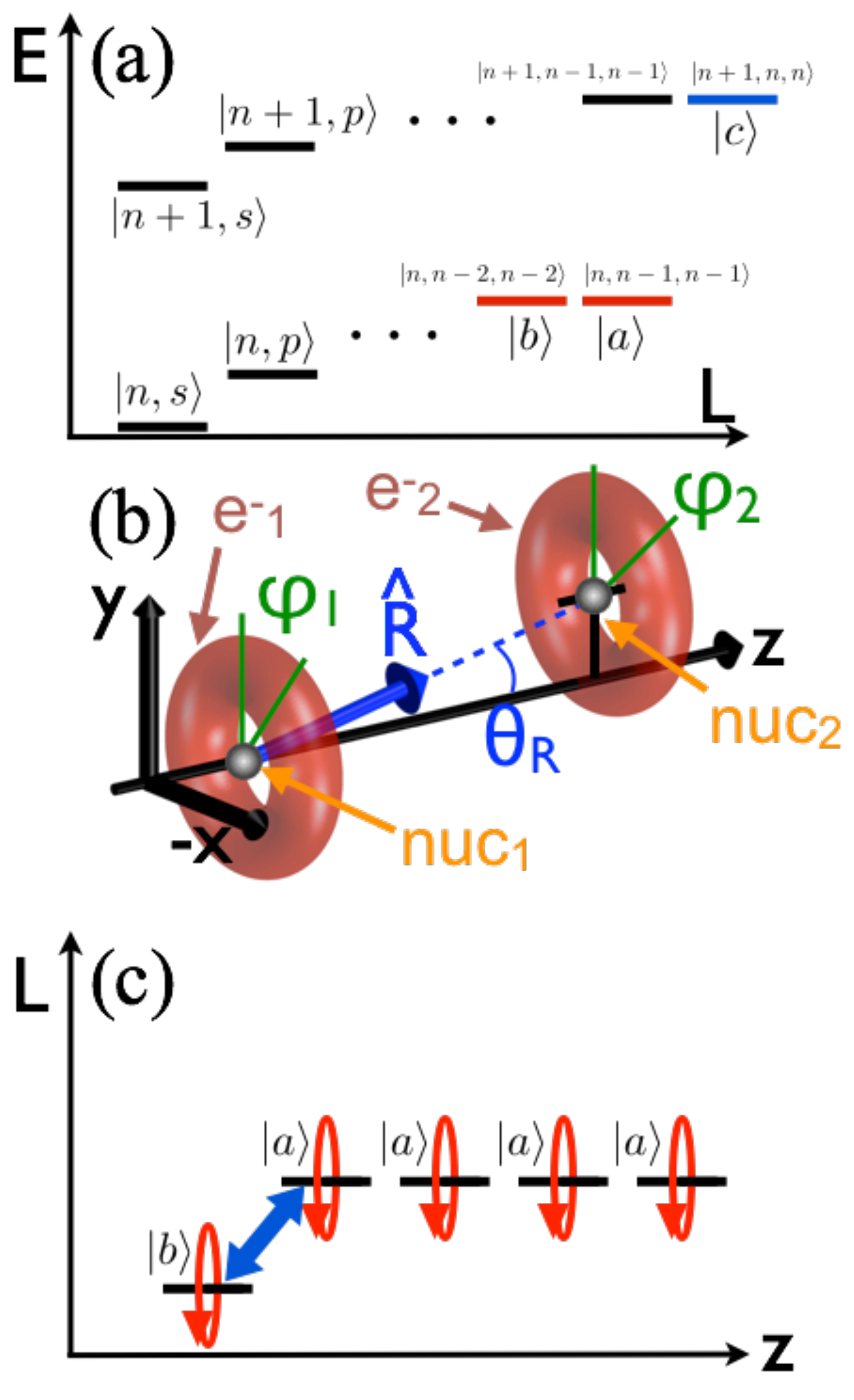,width= 0.8\columnwidth} \\
\caption{ (a) Schematic diagram of energy $E$ versus angular momentum $L$ for low angular momentum- versus high angular momentum Rydberg states. We highlight the special states relevant for this article  $\ket{a}$, $\ket{b}$, $\ket{c}$, defined in the text. (b) Schematic shape of the electron probability distribution (tagged with $e^-_{1,2}$) for two atoms in circular states with angular momentum pointing fully along the quantisation axis $\hat{z}$. Electron orbits are reminiscent of a circular planetary orbit (red toroidal shape). We also indicate nuclear positions and the unit vector along the inter-atomic axis $\hat{\mathbf{R}}$, its angle with the quantisation axis $\theta_R$ and orbital angles for classical electron positions $\varphi_{1,2}$. (c) Controlled angular momentum transport on a chain of Rydberg atoms along the $z$ axis can proceed using only two single atom states, $\ket{a}$, $\ket{b}$, among the high angular momentum manifold.
\label{geometry_states}}
\end{figure}
Rydberg atomic properties are qualitatively changed in circular states, where angular momentum is maximised to $l=n-1$ and pointing along the quantisation axis $m=l$, or nearby $l=n-2$, $m=l=n-2$. Most notably, circular states can have orders of magnitude larger lifetimes than low-l states, ranging into seconds. This has for example been essential in their use for quantum state tomography in cavity quantum electro-dynamics \cite{BrHaLe90_976,brune:processtomog,BrHaRa92_5193,GlKuGu07_297,deleglise:reconstruction} and has recently attracted attention in the context of quantum computing \cite{xia:circgate,saffman:atomic_quant_comp_review} or quantum simulations of spin systems \cite{brune:towardscircular}.
The price paid for the larger lifetime is a substantially more involved excitation process, which has nonetheless been demonstrated also in an ultracold context \cite{anderson_circprod,nussenzveig1993preparation,brecha1993circular,zhelyazkova2016preparation}.

Here we determine the utility of a regular assembly of atoms in circular Rydberg states for studies of excitation- and angular momentum transport as well as a platform for flexible Rydberg aggregates. When working in the quasi-hydrogenic manifold of circular states, the many-body electronic Hilbert space can no longer be conveniently simplified based on energetic separation of undesired states. However dipole-dipole selection rules can still allow simple aggregate state spaces consisting of only the two nearest to circular states listed above, where we will study two choices. These both differ from the electronic states considered in \cite{brune:towardscircular} (in the $n$, $n+2$ manifolds), in that interactions are direct and no two-photon transition is required. We then focus strongly on the implications for exploiting atomic motion.

We theoretically demonstrate clean back and forth transfer of angular momentum
within a Rydberg dimer due to the underlying Rabi oscillations between circular states. We also show that in this regime transport can be described both quantum-mechanically and classically, showing good agreement. Interactions between Rydberg atoms in circular states thus might be a further interesting avenue for studies of the quantum-classical correspondence principle with Rydberg atoms \cite{Gaeta:quasiclassicalstates,hezel:classicalrydberg,hezel:classicalrydbergstark,Samengo:classicalmomryd,Bucher:riseandfall,Deeney:sommerfeld}. Misalignment of the Rydberg aggregate and the electron orbits is shown to cause decreased contrast of the angular momentum oscillations, that can however be suppressed with small electric fields, as also discussed in \cite{brune:towardscircular} for a different choice of states.

We finally explore accessible parameter spaces for Rydberg aggregates based on circular states with the primary focus on flexible Rydberg aggregates (atomic motion), taking into account the main limitations, primarily finite lifetime and adjacent $n$-manifold mixing for too close atomic proximity. We find that flexible aggregates based on circular states offer significantly favorable combinations of lifetime and duration of motional dynamics, despite the weaker interactions, compared to aggregates based on low lying angular momentum states.
The number of aggregate atoms could thus be increased to about $\sub{N}{agg}=50$.

This article is organized as follows: In \sref{circ_states} we introduce circular state atoms and their interactions, leading to a model of excitation transfer on a flexible Rydberg chain. Angular momentum Rabi oscillations in a circular Rydberg dimer are presented in \sref{dimer_transport}, and compared to their classical counterpart. The parameter regimes appropriate for the model in \sref{flexagg} are investigated in \sref{param_regimes}, and then demonstrated in \sref{tully} with an example for angular momentum transport in a large flexible aggregates.

\section{Rydberg atoms in circular states}
\label{circ_states}
Consider an electronic Rydberg state with principal quantum number $n\gg 10$ of an Alkali atom, e.g.~$^{87}$Rb. For a given $n$, we concentrate on the \emph{circular} or almost circular states with the two highest allowed values of angular momentum $l=(n-1),(n-2)$. In both cases, angular momentum shall point as much as possible along the quantisation axis, with azimuthal quantum number $m=+l$.  In the following, we write triplets of quantum numbers $\ket{n,l,m}$ for electronic states of atoms. Then our states of main interest are $\ket{a} = \ket{n,(n-1),(n-1)}$ and $\ket{b}=\ket{n,(n-2),(n-2)}$, the circular and next-to-circular states in the principal quantum number manifold $n$. They can be interpreted in terms of Bohr like orbits, with the electron encircling the nucleus on a circular (or very slightly elliptical) orbit, giving rise to the electron probability densities shown in \frefp{geometry_states}{b} via their isosurfaces, for quantisation axis along $\hat{z}$.

We will additionally consider one third state $\ket{c} = \ket{n+1,n,n}$, the fully circular one in the next higher $n$ manifold, all states are sketched in \frefp{geometry_states}{a}.

\subsection{Effective life times}
\label{taulife}
The change of angular momentum $\Delta l=l_2-l_1$ in a spontaneous electric dipole-transition from state $1$ to state $2$ must fulfill $|\Delta l| =1$, hence circular states must decay towards the ground-state through radiative cascades via the nearest angular momentum state and thus exhibit much longer radiative lifetimes $\tau$ in vacuum than lower angular momentum (Rydberg) states.  At $T=0$ we can use the formula \cite{xia:circgate,beterov:BBR}
\begin{align}
\tau_0&=\frac{24\pi \epsilon_0\hbar^4 c^3}{[E_H^3 a_0^2 e^2]}\frac{(2n-1)^{4n-1}}{2^{4n+1} n^{2n-4} (n-1)^{2n-2}}
\label{tau0live}
\end{align}
for the vacuum lifetime of a circular state in the manifold $n$, which is based on the rate for the first transition of this cascade. In \bref{tau0live} $E_H$ is the Hartree energy and $a_0$ the Bohr radius.
However $\tau_0$ then gets shortened to an effective lifetime $\tau$ by black-body radiation (BBR) at temperature $T$, which accelerates the first step of the cascade by stimulated transitions and may even redistribute electronic population to higher energy states when BBR absorption occurs. We can estimate $\tau$, for $T$ in degrees Kelvin, by
\begin{align}
\tau&=\left( 14.7\mu\mbox{s} \right)\frac{n^2}{T},
\label{taulive}
\end{align}
derived in \cite{cooke_gallagher_BBR1980} by using sum rules. For the state $\ket{53,52,52}$, considered later in \fref{transport} of this article, formula \eref{tau0live} yields a life time of $\tau=38$ ms at $T=0$ but \eref{taulive} an effective lifetime $\tau=138$ ${\mu}$s at $T=300$ K.

\subsection{Binary interactions}
\label{binint}
While long lifetimes are an attractive feature for quantum simulations involving  Rydberg atoms, such simulations typically rely also on a small accessible electronic state space per atom, such that each atom can, for example, be considered as a (pseudo) spin-$1/2$ or spin-$1$ system. This can be realized by Rydberg $\ket{s}$  (l=0) or $\ket{p}$ (l=1) states of the same principal quantum number, provided the energy gap to the $\ket{d}$ state is larger than the dynamical energy scales of the problem, which is frequently the case. In contrast, the high angular momentum states become essentially degenerate approaching Hydrogen states, so simple energetic inaccessibility can no longer be exploited.

However, in principle, interactions can be designed such that still only two circular Rydberg states per atom play a role. This becomes clear by inspection of the dipole-dipole coupling matrix elements, see \aref{appendix_interactions} and e.g.~\cite{book:gallagher,noordam:interactions,arc:package}.
These couple only two-body states with the same total azimuthal quantum number $M=m_1+m_2$, as long as the quantisation axis $\hat{\mathbf{z}}$ is oriented along the inter-atomic separation $\mathbf{R}=\mathbf{x}_2 -\mathbf{x}_1$, where $\mathbf{x}_{1,2}$ are the coordinates of the nucleii in the two interacting atoms. In that case we have $\hat{\mathbf{z}}= \hat{\mathbf{R}}$, where $\hat{\mathbf{R}} = \mathbf{R}/|\mathbf{R}|$.
Dipole-dipole interactions \bref{dipole_hamil} then couple the two pair states $\ket{ab}$, $\ket{ba}$. However, since these are the only pair states with $M=2n-3$ for the principal quantum number $n$ manifold, they form a closed subspace, as long as interactions are weak enough not to cause mixing of adjacent $n$ manifolds.

It is the main objective of this article to explore the limitations of this simple picture. To this end, we consider the more complete Rydberg-Rydberg interactions that arise when taking into account more states and imperfect axis alignment or adjacent $n$-manifold mixing. For this we generate a Rydberg dimer Hamiltonian $\sub{\hat{H}}{pair}$ in matrix form for a fixed atomic separation $\mathbf{R}$ and a large range of pair states $\ket{(n,l,m)_1(n',l',m')_2}$ in the energetic vicinity of those of interest.
In the state notation, $(n,l,m)_k$ are quantum numbers pertaining to atom $k$. Ingredients of the Hamiltonian are all non-interacting pair energies and matrix elements of the dipole-dipole interactions, as discussed in \aref{appendix_interactions}.

We firstly extract dipole-dipole interactions such as $\bra{ba}\sub{\hat{H}}{pair}\ket{ab}\equiv C_3^{(ab)}/R^3$, with $R=|\mathbf{R}|$, see also \aref{appendix_analyticalC3}. Secondly, we determine van-der-Waals interactions in state $\ket{aa}$
by the diagonalization
\begin{align}
\sub{\hat{H}}{pair}(\mathbf{R})\ket{\phi_n(\mathbf{R})}=V_n(\mathbf{R})\ket{\phi_n(\mathbf{R})}.
\label{molpots}
\end{align}
 The interaction potential $V_n(\mathbf{R})$ for which $\ket{\phi_n(\mathbf{R})}\rightarrow \ket{aa}$ for $R\rightarrow\infty$ is then fitted with $V_n(\mathbf{R})\approx C_6^{(aa)}/R^6+V_{n0}$ to infer $C_6^{(aa)}$.

For simplicity, we neglect spin-orbit interactions throughout this article. Their presence will not cause large quantitative or qualitative changes from the conclusions reached here.

\subsection{Many-body interactions in flexible Rydberg aggregates}
\label{flexagg}

Armed with binary interactions inferred as discussed above, we can now reduce the effective electronic state space per atom to include only two states. This then enables us to easily
treat a larger number of atoms.

We consider a multi-atom chain as sketched in \frefp{geometry_states}{(b,c)}, where all atoms are as much as possible aligned with the quantisation axis $\hat{\mathbf{z}}$. While the angle $\theta_R$ between the quantisation axis and inter-nuclear axis $\hat{\mathbf{R}}$ is ideally $\theta_R=0$, we will later consider alignment imperfections $\theta_R\neq0$. In the ideal case, a single "excitation" in the state $\ket{b}$ can migrate through coherent quantum hops on a chain of circular Rydberg atoms in $\ket{a}$, as sketched in \frefp{geometry_states}{(c)}.

Note that creating an initial state such as shown, involving two different circular states poses additional challenges not covered by protocols experimentally demonstrated so far. These only manipulate all atoms in an identical fashion. Possible solutions allowing atom specific manipulation may have to utilise electric field gradients and sequential optical excitation for atom selective addressing and could employ optimal coherent control~\cite{Patsch_circularisation_PhysRevA}.

A setup as in \frefp{geometry_states}{(c)} realizes a Rydberg aggregate \cite{wuester:review}. Since the number of excitations is conserved, we can describe the aggregate in the basis $\ket{\pi_n}=\ket{aa \cdots b \cdots aa}$, where only the $n$'th atom is in the next-to-circular state $\ket{b}$ and all others are in $\ket{a}$. This is called the single-excitation manifold.

The effective electronic Hamiltonian can then be written as
\begin{align}
\sub{\hat{H}}{eff}(\bv{X})&=\sum_{n\neq m}^N \frac{C_3^{(ab)}}{X_{nm}^3} \ket{\pi_n}\bra{\pi_m} + E(\bv{X})\id,
\label{AggHamil}\\
E(\bv{X})&=\frac{1}{2}\sum_{j\ne\ell} \frac{C^{(aa)}_6}{X_{j\ell}^6},
\label{HVdW}
\end{align}
where the vector $\mathbf{X}=[\mathbf{x}_1,\mathbf{x}_2, \mathbf{x}_3\dots]$ groups all the individual positions $\mathbf{x}_n$ of our $N$ atoms, and $X_{nm} = |\bv{x}_n - \bv{x}_m|$, $\id$ is the electronic identity matrix $\id=\sum_n \ket{\pi_n}\bra{\pi_n} $. The first term in \bref{AggHamil} allows excitation transport as discussed above and the second represents van-der-Waals (vdW) interactions between atoms in the $\ket{a}$ state. For simplicity we assumed $C^{(ab)}_6\approx C^{(aa)}_6$. Typically the dipole-dipole interactions dominate vdW in parameter regions where $C^{(ab)}_6\neq C^{(aa)}_6$ would make a difference, however see \cite{zoubi:VdWagg} for counter examples.

To describe a flexible aggregate with mobile atoms we solve
\begin{equation}
\sub{\hat{H}}{eff}(\bv{X})\ket{\varphi_n(\bv{X})} = U_n(\bv{X})\ket{\varphi_n(\bv{X})}
\label{eigensystem}
\end{equation}
and obtain the excitonic Born-Oppenheimer surfaces $U_n(\bv{X})$ that govern the atomic motion, see \cite{wuester:review}.

\section{Rydberg dimer with circular states}
\label{dimer_transport}
We begin to study angular momentum transport between a pair of Rydberg atoms in circular states for a simple dimer shown in \frefp{geometry_states}{b}. This allows us to still use the Hamiltonian $\sub{\hat{H}}{pair}$ based on a larger number of electronic states per atom. We employ the time-dependent Schr{\"o}dinger equation (TDSE) $i\hbar \frac{\partial}{\partial t} \ket{\Psi} = \sub{\hat{H}}{pair}  \ket{\Psi}$, where the Hamiltonian is constructed as discussed in \sref{circ_states} and \aref{appendix_interactions}. Within that space
\begin{align}
\ket{\Psi(t)} &= \sum_{nlm,n'l'm'}  c_{nlm,n'l'm'}(t) \ket{(nlm)_1(n'l'm')_2},
\label{resspace_TDSE}
\end{align}
where $(nlm)_1$ are quantum numbers of atom $1$.

The dimer is initialized in the pair state $\ket{\Psi(0)}=\ket{ab}$ for the $n=53$ manifold. As discussed in \sref{circ_states}, dipole-dipole interactions cause transitions to the pair state $\ket{ba}$, giving rise to Rabi-oscillations in an effective two level system, shown in \frefp{transport}{a}. For now, the inter-atomic axis is perfectly aligned with the quantisation axis ($\theta_R=0$). Physically this implies that Rydberg electron orbitals are orthogonal to the interatomic axis. The figure shows  the modulus of electronic angular momentum per atom $\langle \hat{L}_1\rangle = \sum_{nlm,n'l'm'} \hbar \sqrt{l (l+1)}|c_{nlm,n'l'm'}|^2$, $ \langle \hat{L}_2\rangle = \sum_{nlm,n'l'm'} \hbar \sqrt{l' (l'+1)}|c_{nlm,n'l'm'}|^2$.

\subsection{Quantum classical correspondence}
\label{classical_dimer}
The angular momentum exchange can also be modeled classically, using Newton's equation for the Rydberg electrons, with results shown in black in \frefp{transport}{a}. Further details of these simulations can be found in \aref{appendix_classical}. Already the simple model employed reproduces the quantum results almost quantitatively. This is expected for circular Rydberg states, since the number of de-Broglie wavelengths $\sub{\lambda}{dB}$ fitting into one orbital radius $\sub{r}{orb}$ equals $\sub{r}{orb}/\sub{\lambda}{dB} = n$ in Bohr-Sommerfeld theory, which reduces the importance of quantum effects (wave features) for large $n$, in accordance with the correspondence principle.

The result indicates the utility of interactions among circular Rydberg atoms to illustrate the correspondence principle in action. Once verified in more detail, classical simulations could then supplement quantum ones in the regime where each atom accesses a large number of electronic states, which are challenging quantum mechanically.

\begin{figure}[htb]
\centering
\epsfig{file=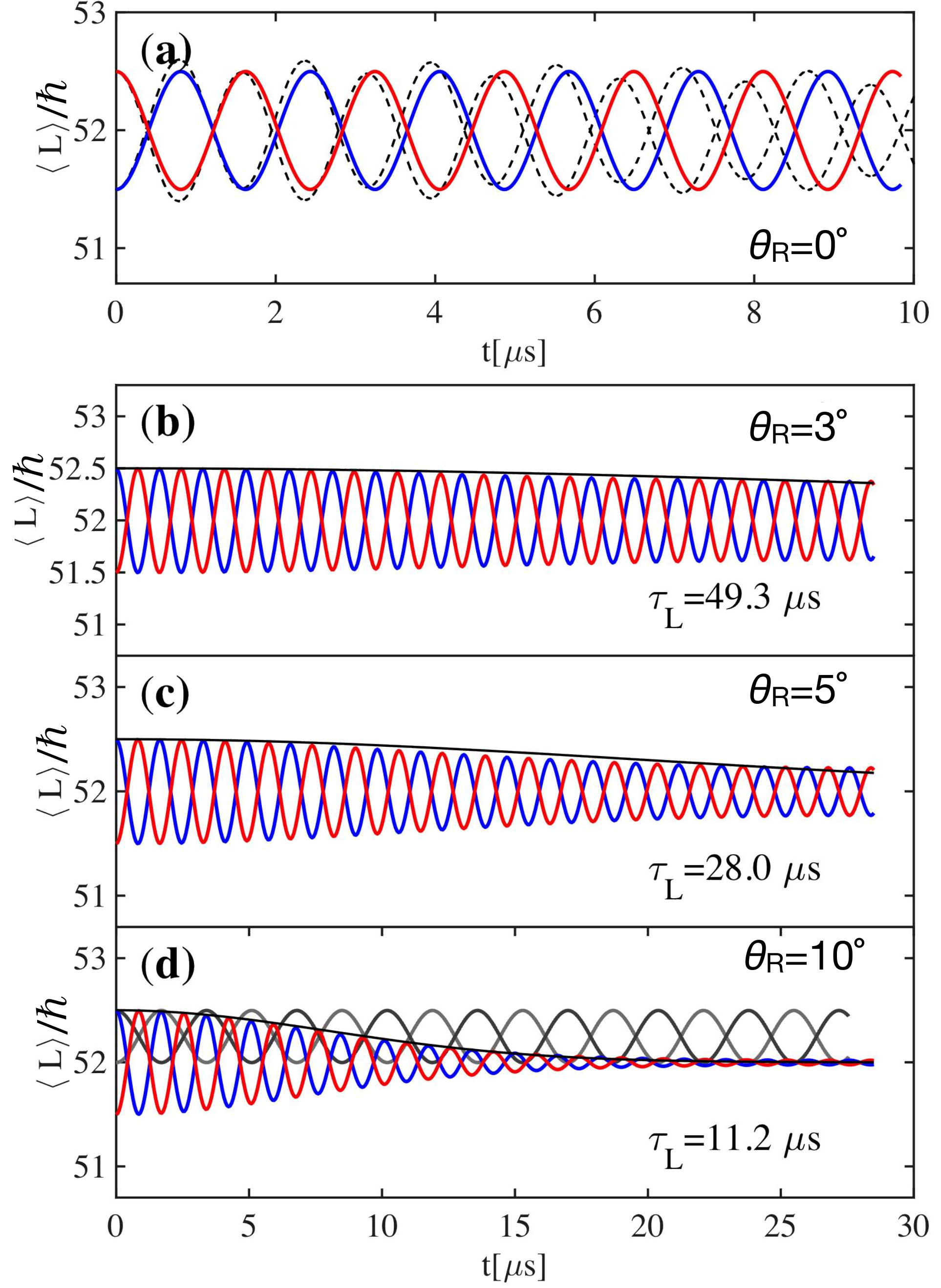,width= 0.99\columnwidth}
\caption{Angular momentum transport in a dimer of $n=53$ circular state Rydberg atoms, separated by $R=10$ $\mu$m, after initialisation in $\ket{\Psi(0)}=\ket{ab}$. Solid lines show the quantum mechanical results for the angular momentum one each atom $\expec{\hat{L}_1}$ (red, starting at $\expec{\hat{L}_1}=52.5$) and $\expec{\hat{L}_2}$ (blue, starting at $\expec{\hat{L}_2}=51.5$). Black dashed lines are the corresponding angular momenta from the classical Newton's equations, see \aref{appendix_classical}. In (a), electron orbital planes for state $\ket{a}$ are perfectly normal to the inter-atomic axis. The angle $\theta_R$ between $\hat{R}$ and $\mathbf{z}$ (see \frefp{geometry_states}{b}) is $\theta_R=0^{o}$. (b) Quantum mechanical angular momenta for a misaligned dimer with $\theta_R=3^o$. The solid line is a fit on the envelope as discussed in the text. (c) $\theta_R=5^o$. (d) $\theta_R=10^o$. The grey lines without reduction of oscillation amplitude show the corresponding result in the presence of a small electric field, see text.
\label{transport}}
\end{figure}
%

\subsection{Misalignment of electron orbits and inter-atomic separation}
\label{misalignment}
In the remainder of \fref{transport}, we explore how a misalignment of the circular orbits from the interatomic axis, $\theta_R>0$, affects angular momentum transport. For that case $M=m_1+m_2$ is no longer conserved in dipole-dipole interactions (see \aref{appendix_interactions}). Hence a large number of different azimuthal states $m\neq \{n-1,n-2\}$ become populated. This brings into play additional dipole-dipole interaction matrix elements that cause angular momentum transfer between the two atoms.
Since these differ in magnitude, the overall angular momentum oscillations in $L_{1,2}$ loose contrast as seen in \frefp{transport}{(b)}-(d).  We fitted the envelope of oscillations with $\exp{[- t^2/\tau_L^2]}$ and indicated the resultant $\tau_L$ in the figures.

Note, that even a relative large misalignment such as $\theta=5^o$ still allows many visible periods of angular momentum oscillations. The coupling to undesired azimuthal state can however be entirely suppressed by the addition of an electric field. This removes the degeneracy of different $|m|$ states through the dc Stark effect \cite{book:gallagher}. For \frefp{transport}{d} we used an electric field amplitude ${\cal E}=0.2$ V/cm and initialised the dimer in $\ket{\Psi(0)}=\ket{(53,52,52)_1}\otimes(\ket{(53,51,51)_2} + \ket{(53,52,51)_2})/\sqrt{2}\equiv \ket{a\tilde{b}}$. Note that $\ket{\tilde{b}}=(\ket{53,51,51} + \ket{53,52,51})/\sqrt{2}$ is the Stark coupled eigenstate corresponding to $\ket{b}$ in the presence of the field. While the Rabi frequency is now reduced by a factor of two, since the dipole-dipole interaction couples only the first component of $\ket{\tilde{b}}$ to $\ket{a}$, we regain an effective two-level system. Calculations with electric field where streamlined by solving the TDSE only in the most relevant statespace \cite{footnote:space_reduction}. Suppressing coupling to undesired $m$-states through an external field was explored in detail in \cite{brune:towardscircular} for coupled states from \emph{different} (next-to-adjacent) n-manifolds. Here we now extended these concepts to almost circular states from the same $n$-manifold.

\subsection{Adjacent $n$-manifold mixing}
\label{mixing}
So far, we explored one limitation of the simple picture in which only circular states $\ket{ab}$ and $\ket{ba}$ are considered, namely undesired $m$-levels mixing in through atomic misalignment. We have shown that this effect can be suppressed
 using external electric fields.

\begin{figure}[htb]
\centering\epsfig{file=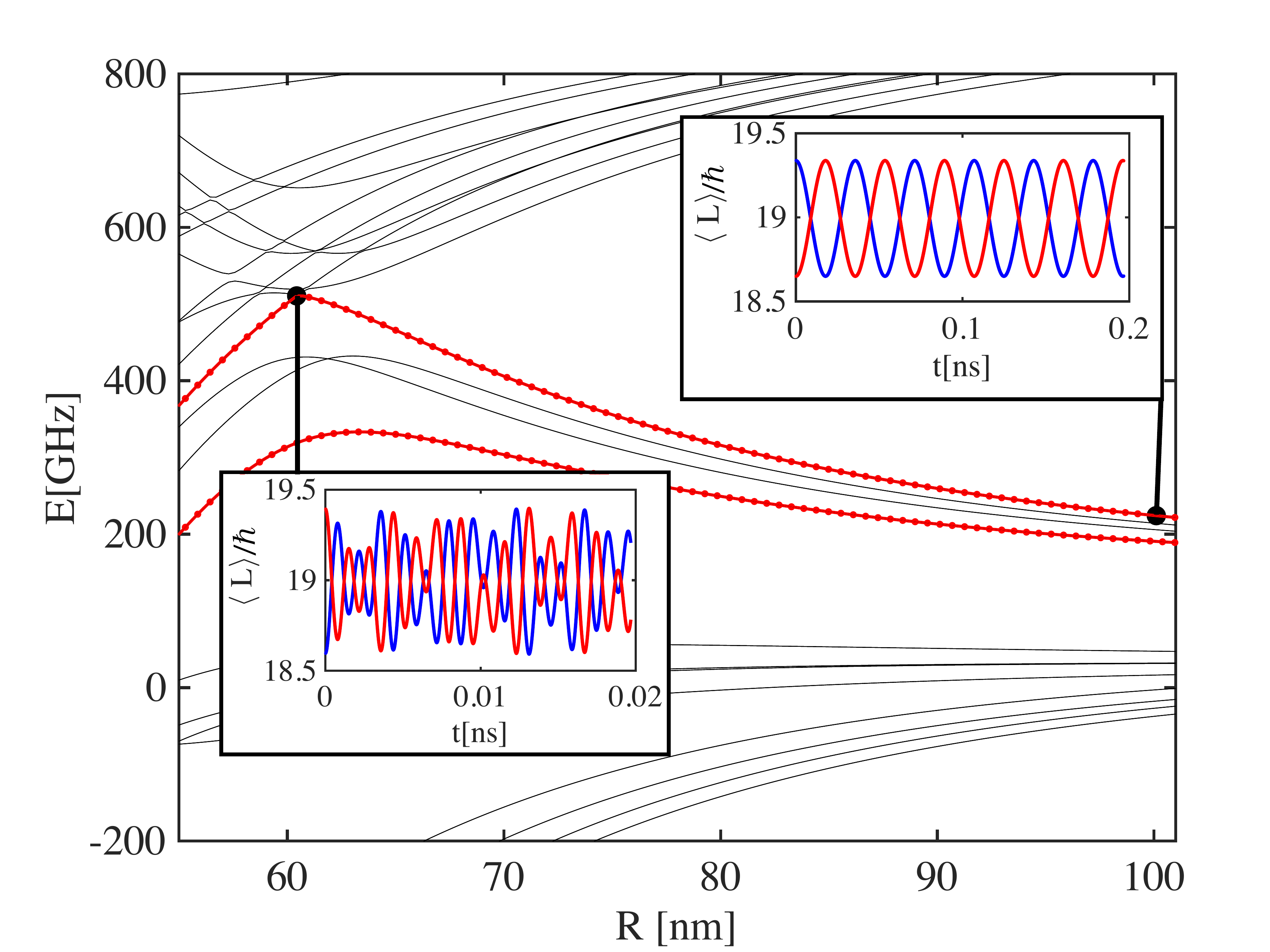,width= 0.99\columnwidth}
\caption{Interaction potentials $V_n(R)$ of a circular Rydberg dimer near $n=20$ at close proximity, see \eref{molpots}. The reduced Hilbert-space contained all states with $n=19$, $20$, $21$  and $l=18$, $19$, $20$. The simple effective state picture involving only two circular states $\ket{a}$ and $\ket{b}$, that couple via dipole-dipole interactions to $(\ket{ab}\pm\ket{ba})/\sqrt{2}$ (red lines with dots), breaks down once neighboring $n$ manifolds begin to merge into each other around $R=60$ nm. The insets show angular momentum transport from initial states as in \frefp{transport}{a} at the indicated separations.
\label{spaghetti}}
\end{figure}

Another limitation of the simple model arises at too short distances, where state manifolds that differ in principal quantum number $n$ are shifted into each other through strong interactions. We show the resultant spectrum in \fref{spaghetti}, for a much lower principal quantum number ($n=20$) than used in \fref{transport}, due to computational reasons.

For demonstration, the figure also shows the detrimental effect on angular momentum transport through this state mixing.
The right inset shows angular momentum oscillations that are regular at distances where adjacent $n$-manifolds are energetically separate. However even here the initial state is composed of eigenstates from \bref{molpots} according to $(\ket{\phi_{ab}(\mathbf{R})} + \ket{\phi_{ba}(\mathbf{R})})/\sqrt{2}$, where $\ket{\phi_{ab}}$ denotes the eigenstate of $\sub{\hat{H}}{pair}$ that has the largest overlap with $\ket{ab}$.
Oscillations finally become irregular at separations where adjacent $n$-manifolds mix, shown in the left inset, even when constructing an initial state from four eigenstates similar to the construction above.
This effect imposes a minimal separation $\sub{d}{min}$ for atoms in a circular Rydberg aggregate, which we define as the distance at which the dipole-dipole shift exceeds the energetic $n$-manifold separation. The resultant formula is given in \aref{appendix_constraints}.

\section{Parameter regimes for circular Rydberg aggregates}
\label{param_regimes}
After exploring the limitations of the simple model introduced in \sref{flexagg}, which are not problematic for the right choice of atomic positions $\mathbf{x}_n$, we now proceed to determine interaction parameters required for the model \bref{AggHamil} as discussed in \sref{binint}.

\subsection{Determination of interaction constants}
\label{interaction_constants}
For dipole-dipole interactions we extract the matrix elements $\bra{ab}\sub{\hat{H}}{pair}\ket{ba}$ and $\bra{ac}\sub{\hat{H}}{pair}\ket{ca}$ from the numerical Hamiltonian, and verify the former analytically in \aref{appendix_analyticalC3}.
Nextly we consider van-der-Waals (vdW) interactions for two atoms in the state $\ket{a}$ (i.e.~the energy of $\ket{aa}$). We find these by diagonalising a suitable Hamiltonian as a function of atomic separation $R$, as discussed in \sref{binint} and \aref{appendix_interactions}. All these calculations assume an internuclear axis aligned with the quantisation axis $\hat{\mathbf{R}}\parallel\hat{\mathbf{z}}$, which is enough to determine the scale of interactions in a setting such as \frefp{geometry_states}{c}.

All interactions exhibit a characteristic scaling with principal quantum number $n$:
\begin{align}
C_6^{(aa)}&=\tilde{C}^{(0)}_6 n^{12}, \label{VVdw}  \\
C_3^{(ab)}&=\tilde{C}^{(0)}_{3,ab} n^{3}\hspace{0.3cm}\mbox{ for }\ket{ab}, \label{Vdd} \\
C_3^{(ac)}&=\tilde{C}^{(0)}_{3,ac} n^{4}\hspace{0.3cm}\mbox{ for }\ket{ac},  \label{Vddac}
\end{align}
which allows their approximate representation in terms of the reference values $\tilde{C}^{(0)}_k$ given in \tref{interaction_parameters}. The table distinguishes between dipole-dipole interactions within the same or among adjacent n-manifolds. Note, that the scaling of interactions with $n$ is different from that encountered for low-lying angular momentum states, where dipole-dipole interactions scale as $n^4$ and van-der-Waals ones as $n^{11}$ \cite{book:gallagher}. VdW interaction strengths from \eref{VVdw} and \tref{interaction_parameters} for circular states with $n=48$ and $n=50$ are in rough agreement with the values given in \cite{brune:towardscircular}, the latter calculated at non-zero electric and magnetic fields.
\begin{table}
\begin{center}
\begin{tabular}{|c|c|c|}
\cline{1-3}
\Big.
      & $\tilde{C}^{(0)}_3$[kHz $\mu m^3$]      &  $\tilde{C}^{(0)}_6$[Hz $\mu m^6$]        \\
\cline{1-3}
$\tabvspace \ket{aa}$  &      & ${ 2.11\times 10^{-11}}$    \\
$\tabvspace  \ket{ab}$   & ${2.0}$   & \\
$\tabvspace \ket{ac} $   & ${0.47}$  &   \\
\cline{1-3}
\end{tabular}
\end{center}
\caption{Reference values in interaction parameters for dipole-dipole and van-der-Waals interactions of $^{87}$Rb atoms in circular or next-to-circular Rydberg states. Using these parameters, interaction strengths can be found with \eref{VVdw}-\bref{Vddac}. States $\ket{a}$, $\ket{b}$, $\ket{c}$ are sketched in \fref{geometry_states} and defined in \sref{circ_states}.
\label{interaction_parameters}}
\end{table}
%

\subsection{Domains for Flexible Rydberg aggregates}
\label{parameeter_domains}
Dipole-dipole interactions in the pair $\ket{ac}$ substantially exceed those in $\ket{ab}$ for the relevant high principal quantum numbers ($n>20$), due to the steeper scaling in $n$. We thus now assume aggregates based on
states $\ket{\tilde{\pi}_n}=\ket{aa \cdots c \cdots aa}$, where only the $n$'th atom is in the state $\ket{c}$ and all others in $\ket{a}$, replacing the states $\ket{\pi_n}$ in \sref{flexagg}.

With interactions determined, we can follow the approach taken in \cite{wuester:review}, to delineate parameter regimes in which circular flexible or static Rydberg aggregates are viable, based on a variety of requirements that are listed in detail in \aref{appendix_constraints}. The results are shown in \fref{circagg_regimes}. It is clear that the use of circular Rydberg atoms for studies involving atomic motion offers substantial advantages. However, this is the case only in a cryogenic environment at $T\approx 4$ K, since black-body redistribution has a too detrimental effect on the lifetime advantage otherwise. Ideas to suppress spontaneous decay by tuning the electro-magnetic mode structure with a capacitor could improve this situation further \cite{Kleppner_inhibspont,hulet_inhibspontexp,brune:towardscircular}.
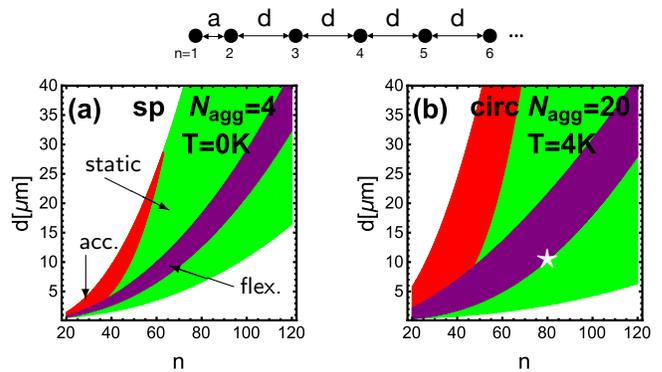
\begin{figure}[htb]
\centering
\vspace{1cm}
\begin{overpic}[width=0.99\columnwidth,scale=1.0,unit=1mm]{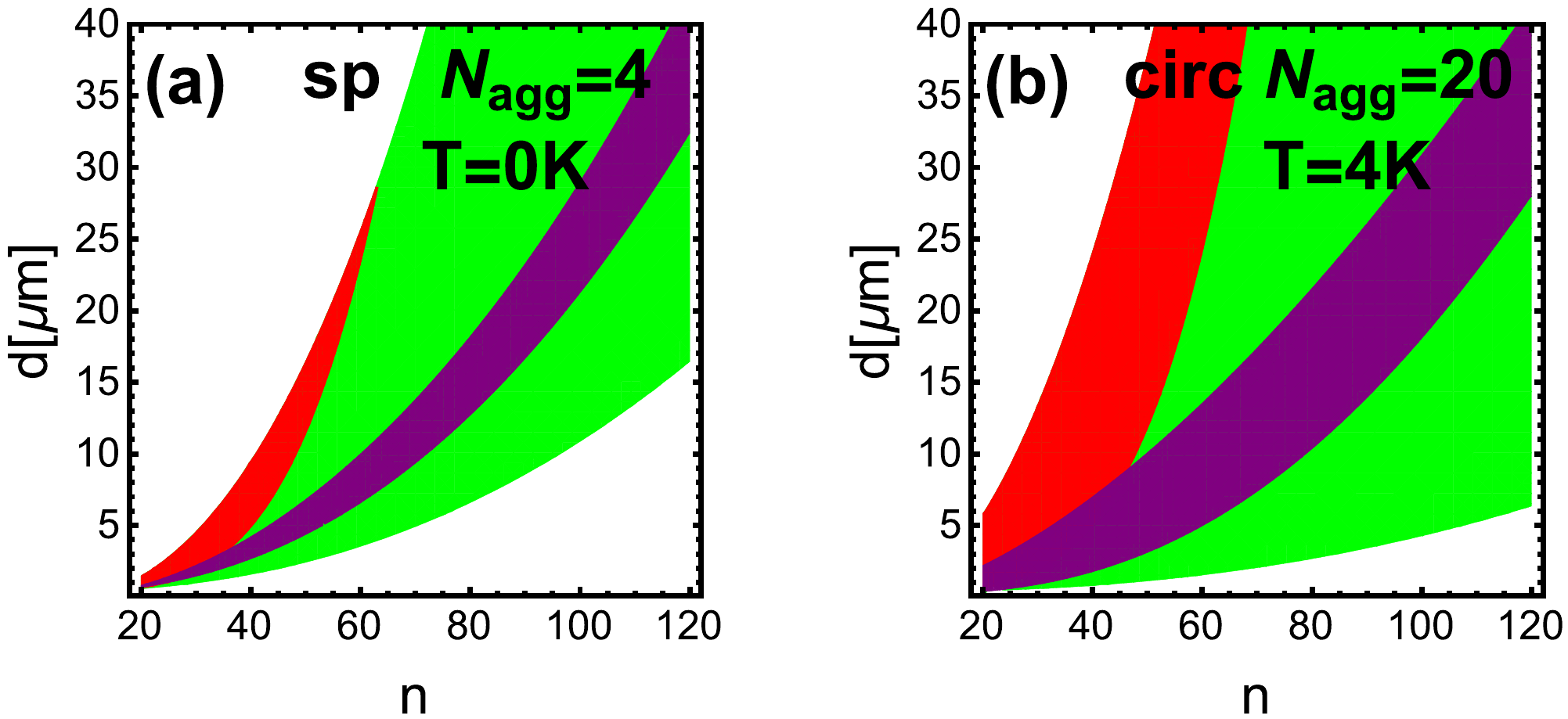}
    \put(82,16){\Large\color{white}$\boldsymbol{\star}$}
    \put(25,48){\includegraphics[width=0.55\columnwidth]{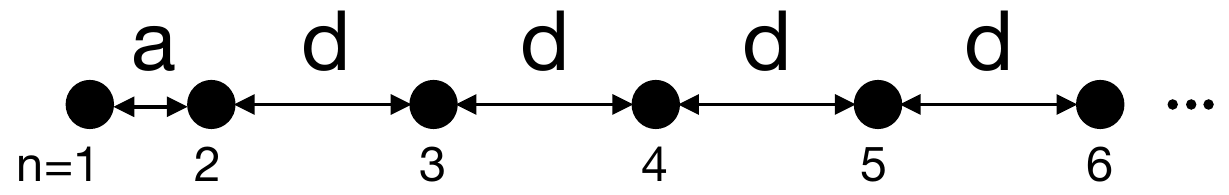}}
    \put(35,13){\color{black}\vector(-3,1){10}}
    \put(36,12){\sffamily flex.}
    \put(15,30){\color{black}\vector(2,-1){10}}
    \put(12,31){\sffamily static}
    \put(12,18){\color{black}\vector(0,-1){7}}
    \put(11,19){\sffamily acc.}
\end{overpic}
\caption{
Parameter domains of static (green and red) versus flexible (violet) Rydberg aggregates, for different principal quantum numbers $n$ and nearest neighbor separations $d$. For the latter we assume geometry as shown on the top, with main nearest neighbor separation $d$ and shorter initial dislocation $a$.
We compare the use of $sp$ Rydberg states in (a) versus circular Rydberg states in (b), where the latter are assumed to be in a cryogenic environment at $T=4K$. Note the substantially different aggregate sizes  $\sub{N}{agg}$ assumed for either as indicated. Red shade (marked acc.) indicates where static aggregates exist, however with atoms that would visibly accelerate during excitation transport. White areas are excluded, either by too short aggregate lifetimes (top) or too close proximities to avoid Rydberg state mixing as in \fref{spaghetti} (bottom). See the text and \aref{appendix_constraints} for the precise criteria used.  The symbol ($\star$) in (b) indicates parameters used for our numerical demonstration in \fref{flexible_circagg_demonstration}.
\label{circagg_regimes}}
\end{figure}
%

\section{Angular-momentum transport in large flexible Rydberg aggregates}
\label{tully}

To illustrate the potential of circular state Rydberg aggregates for studying the coupling between atomic motion and excitation transport, we show
a quantum classical simulation of adiabatic excitation transport on a large ($\sub{N}{agg}=20$) Rydberg aggregate. Adiabatic excitation transport in Rydberg aggregates was thoroughly discussed in \cite{wuester:cradle,moebius:cradle,leonhardt:orthogonal}. Briefly: a single excited state is initially coherently shared among two atoms at one end of the chain, that are in much closer proximity $a$ than all others, here $a=5$ $\mu$m. These are atoms $n=1,2$ in the sketch on top of \fref{circagg_regimes}. This initial state, $\ket{\sub{\varphi}{rep}}=(\ket{caaa\dots}+\ket{acaa\dots})/\sqrt{2}$, is the most repulsive eigenstate in \eref{eigensystem}.

The initial repulsion of atoms $1$ and $2$ causes subsequent repulsive collisions with the remainder of the atoms, the dislocation thus propagating through the chain. The single excitation is carried along with the positional dislocation with high fidelity. This can be traced back to an adiabatic following of the initial dipole-dipole eigenstate $\ket{\sub{\varphi}{rep}(\bv{X}(t))}$, see \eref{eigensystem}.

We model the process using Tully's surface hopping \cite{tully:original,tully:original2,Barbatti:tully_review}, described for our specific purposes in \cite{leonhardt:thesis,leonhardt:orthogonal}.
It evolves an electronic aggregate quantum state $\ket{\sub{\Psi}{agg}(t)}=\sum_n c_n(t) \ket{\pi_n}$, coupled to the classical Newton equations $\sub{m}{Rb}\ddot{\mathbf{X}}(t)=-\boldsymbol{\nabla}_\mathbf{X} U_s(\mathbf{X}(t))$ for motion of Rubidium atoms with mass $\sub{m}{Rb}$ on the current Born-Oppenheimer surface $U_{s(t)}$, see \eref{eigensystem}. Note, that creating the initial electronic state $\ket{\sub{\varphi}{rep}}$ will pose additional experimental challenges.

The parameters used for the simulation are indicated by the white star in \fref{circagg_regimes}, and the (one-dimensional) geometry sketched on top of that figure. For these parameters, even $\sub{N}{agg}=100$ would still allow end-to-end transport within the lifetime, however with long simulation times due to the need for matrix diagonalisation at each time-step.

\begin{figure}[htb]
\centering
\epsfig{file=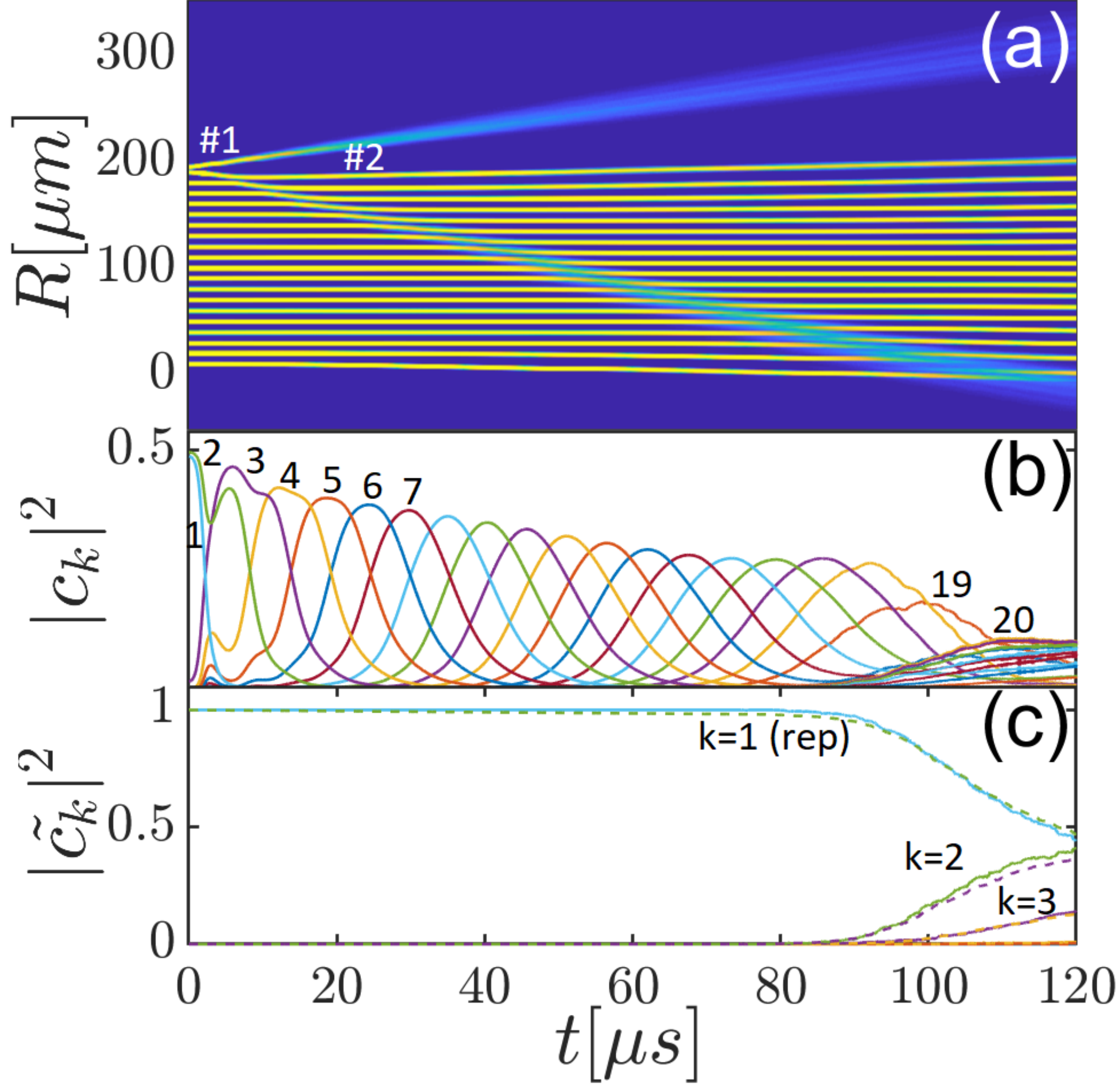,width= 0.99\columnwidth}
\caption{Adiabatic angular momentum transport on a large flexible Rydberg aggregate with $N=20$ atoms arranged in a one-dimensional line along $z$ with spacing $d=10$ $\mu$m, but last two atoms only $a=5$ $\mu$m apart. Dynamics proceeds on the repulsive Born-Oppenheimer surface $n=0$. The aggregate is based on circular states $\ket{a}$, $\ket{c}$ with principal quantum number $n=80$. For that value, the effective life-time from \eref{taulive} for the \emph{entire aggregate} is $\sub{\tau}{agg}=\tau/\sub{N}{agg}\approx 1.2$ ms at $T=4$K.
Each atom has a spatial position uncertainty of $\sigma = 0.3$ $\mu$m.
(a) Total density of atoms, bright/yellow indicates high density, blue/dark no density. (b) Excitation amplitudes on each atom $|c_k|^2 = |\braket{\pi_k}{\Psi(t)}|^2$, with atom number $k$ indicated near each line. We indicate where numbering starts in (a). (c) Populations of system eigenstates $|\tilde{c}_k|^2=|\braket{\varphi_k(\bv{X})}{\Psi(t)}|^2$, discussed in \sref{flexagg}, indicating largely adiabatic dynamics.
\label{flexible_circagg_demonstration}}
\end{figure}

Proposals in \cite{wuester:cradle,moebius:cradle,leonhardt:orthogonal} were limited by spontaneous decay to about eight Rydberg atoms, even when considering the lighter, and thus more easily accelerated, Lithium atom. The quantum-classical simulation shown in \fref{flexible_circagg_demonstration} highlights that for aggregates made of circular states much larger arrays are possible even for the heavier but more common Rubidium atom, and still show adiabatic excitation transport within the system life-time, i.e.~well before a single black-body redistribution event is expected.

While the multi-trajectory average in \frefp{flexible_circagg_demonstration}{b} seems to indicate a loss of fidelity for the excitation transport, this is merely due to the different arrival times for different parts of the many-body wavepacket (different trajectories). We inspected many individual quantum-classical trajectories, which all show near unit fidelity of excitation transport through the entire chain.

\section{Conclusions and Outlook}
\ma{\label{conc}}

We assess the utility of arrays of Rydberg atoms in circular and nearly circular angular momentum states for the realisation of flexible Rydberg aggregates.
While the motion of circular state Rydberg atoms was considered in \cite{brune:towardscircular} as a precursory stage during the creation of regular \emph{static} arrays,
in our work freely moving atoms are the primary focus. These will then allow studying the inter-relationship between atomic motion and excitation or angular momentum transport.
Note that the apparatus proposed in \cite{brune:towardscircular} would also be highly suitable for such studies.

In a cryogenic environment (suppressing black-body radiation), circular state flexible Rydberg aggregates will allow much larger arrays of atoms to participate in collective motional dynamics despite their inertia, due to the substantially increased lifetimes. For example, adiabatic excitation transport with high fidelity on chains of as many as $\sub{N}{agg}=50$ atoms appears feasible. In the future we will explore the application of this phenomenon for use as a data-bus in circular Rydberg atom based quantum computing architectures \cite{xia:circgate,saffman:atomic_quant_comp_review}.

We also demonstrate a case where interacting circular Rydberg atoms can be quite well described using the classical Newton's equations for the Rydberg electrons in a manifestation of the correspondence principle. Both quantum and classical calculations exhibit comparable coherent angular momentum oscillations in a pair of circular Rydberg atoms. More detailed comparisons using more involved classical phase space distributions and quantum wave packets, larger numbers of atoms or more involved geometries could be an interesting exploration of the extent of the correspondence principle. A classical treatment of interactions could then benefit from secular perturbation theory techniques also used in planetary orbital mechanics.

\appendix
\section{Circular Rydberg interactions}
\label{appendix_interactions}

We assume the inter-atomic interactions are entirely based on the dipole-dipole component of the electro-static Hamiltonian (in atomic units)
\begin{align}
\hat{H}_{dd}=\frac{\bv{r}_{1}\cdot \bv{r}_{2} - 3 (\bv{r}_{1}\cdot\hat{\bv{R}}) (\bv{r}_{2}\cdot\hat{\bv{R}})}{R^3},
\label{dipole_hamil}
\end{align}
where $\bv{r}_{1,2}$ denote the position of the Rydberg electron in atom $1,2$ relative to their parent nucleii, and $\hat{\bv{R}}=\mathbf{R}/R$ is a unit vector along the inter-atomic separation $\mathbf{R}=\mathbf{x}_2-\mathbf{x}_1$, with $R=|\mathbf{R}|$, see \frefp{geometry_states}{b}.
We thus ignore wave-function overlap, core-polarisation or higher order multipoles, as is typical for Rydberg-Rydberg interactions.

We then cast \bref{dipole_hamil} into a matrix form using pair states $\ket{n_1,l_1,m_1}_1\otimes\ket{n_2,l_2,m_2}_2$ in a truncated Hilbertspace, in which all pair-states are energetically close to those for which we want to determine Rydberg-Rydberg interactions.
As usual, the position space representation is written as $\braket{\mathbf{r}_1}{n_1,l_1,m_1}={\cal R}_{n_1l_1}(r_1) Y_{l_1,m_1}(\theta_1,\varphi_1)/r_1$, where $Y$ are spherical harmonics, and $(r_1,\theta_1,\varphi_1)$ the 3D spherical polar coordinates of electron one with respect to its nucleus.

Matrix elements of \bref{dipole_hamil} are
\begin{align}
&\langle n_1,l_1,m_1;n_2,l_2,m_2 |\hat{H}_{dd} |n'_1,l'_1,m'_1;n'_2,l'_2,m'_2 \rangle
\CR
&=-8\pi \sqrt{\frac{2\pi}{15}}\frac{d_{n_1,l_1;n'_1,l'_1} d_{n_2,l_2;n'_2,l'_2}}{R^3}
\CR
&\times\sum_{m_a,m_b} \sum_{\mu=-2}^{2}Y_{l=2,\mu}^*(\theta_R,\varphi_R) \braket{1m_1,1m_2}{2\mu}
\CR
&\times \bra{l_1,m_1}Y_{1m_1} \ket{l'_1,m'_1}\bra{l_2,m_2}Y_{1m_2}\ket{l'_2,m'_2},
\label{dipole_hamilMEs}
\end{align}
see also  \cite{noordam:interactions}. Here $\theta_R$, $\varphi_R$ are the polar angles of $\hat{\mathbf{R}}$ in the 3D spherical coordinate system defining $n,l,m$, $\braket{1m_1,1m_2}{2\mu}$ the Clebsch-Gordan coefficient coupling two constituent angular momenta $(l=1,m=m_{1,2})$ to a total angular momentum $(L=2, M=\mu)$ and the integrals in the last line involve now a single electronic co-ordinate and three spherical harmonics each.

Evaluating these as in \cite{book:gradshteyn}, we use
\begin{align}
\bra{l_1,m_1}Y_{1m_1}&\ket{l'_1,m'_1}= (-1)^{m_1} \sqrt{\frac{3(2l_1+1)(2l'_1+1)}{4\pi}}
\CR
\times
&\left(\begin{array}{ccc}
l_1&  l'_1 & 1 \\
0 & 0 & 0 \\
\end{array}\right)
\left(\begin{array}{ccc}
l'_1&  l_1 & 1 \\
m'_1& -m_1 & m_1 \\
\end{array}\right),
\label{dipoleME_singat}
\end{align}
where terms in brackets denote Wigner $3j$ symbols.

The $d_{n_1,l_1;n'_1,l'_1}=\int_0^\infty r \: {\cal R}_{n_1,l_1}(r) {\cal R}_{n'_1,l'_1}(r) dr$ in \bref{dipole_hamilMEs} are radial matrix elements, determined via the Numerov method including modifications of the Coulomb potential due to the core as in \cite{amthor:thesis}. To avoid instabilities, the solutions ${\cal R}(r)$ are set to zero inside the \emph{inner} classical turning point for large $l$.

When considering interactions within an external electric field of strength ${\cal E}$, we describe the field through single body matrix elements
\begin{align}
&-\langle n,l,m| {\cal E}e \hat{\bv{z}} |n',l',m' \rangle
=-d_{n,l;n',l'}{\cal E}e
\CR
\times
&\sqrt{\frac{3(2l+1)(2l'+1)}{4\pi}}
\left(\begin{array}{ccc}
l&  l' & 1 \\
0 & 0 & 0 \\
\end{array}\right)
\left(\begin{array}{ccc}
l'&  l & 1 \\
m'& -m & 0 \\
\end{array}\right).
\label{Efield_hamil}
\end{align}
To obtain vdW interaction potentials, the resultant dimer Hamiltonian $\sub{\hat{H}}{pair}=\hat{H}_0 + \hat{H}_{dd}$ is diagonalized as a function of separation $R$, see \bref{molpots} and e.g.~\fref{spaghetti}. Here, the non-interacting Hamiltonian is $\hat{H}_0 = \sum_{\boldsymbol{\alpha}_1,\boldsymbol{\alpha}_2} (E_{\boldsymbol{\alpha}_1} + E_{\boldsymbol{\alpha}_2}) \ket{\boldsymbol{\alpha}_1\boldsymbol{\alpha}_2}\bra{\boldsymbol{\alpha}_1\boldsymbol{\alpha}_2}$, where the $\boldsymbol{\alpha}$ group all electronic labels, such as~$\boldsymbol{\alpha}_1=\{n,l,m\}$ with $E_{\boldsymbol{\alpha}_1}=E_{n_1,l_1,m_1} = -Ry/(n_1 - \delta_{n_1,l_1})^2$. Then $Ry$ is the Rydberg constant and $\delta_{n,l}$ the quantum defect taken from \cite{amthor:thesis,Li_Gallagher_Rbquantdefects2003}. For transport simulations, the restricted basis Hamiltonian is constructed at a fixed separation $R_0$ and then used in the time-dependent Schr{\"o}dinger equation.

A recent numerical package for these sort of calculations is described in \cite{weber:rydint:tutorial}. For low lying state interaction, also perturbation theory can be used \cite{singer:C6numbers}.

\section{Calculation of dipole-dipole interaction constants}
\label{appendix_analyticalC3}

For circular states of Alkali atoms, the wave function overlap with the core becomes so small that the use of Hydrogen wave functions $\Psi(r_k,\theta_k,\varphi_k)={\cal R}_{nl}(r_k)Y_{lm}(\theta_k,\varphi_k)/r_k$, where $k\in\{1,2 \}$ numbers the atom, becomes highly justified.
We can then determine e.g.~$C_3^{(ab)}$ coefficients from \eref{dipole_hamil} by inserting the appropriate sets of quantum numbers into the matrix element ${\cal M}= \bra{ab}\sub{\hat{H}}{dd}\ket{ba}$ between Hydrogen states.

Since $\hat{\mathbf{R}}\parallel\hat{\mathbf{z}}$ we have $\theta_R=0$. In that case, only $Y_{l=2,0}^*(\theta_R=0,\varphi)$ in the sum over $\mu$ is nonzero, and out of the options for $m_1+m_1'=0$
only one set fulfills the remaining angular momentum selection rules in \bref{dipole_hamilMEs}, yielding the integral
\begin{align}
I=& \braket{1,1;1,-1}{2,0}  \bra{a}Y_{1,1} \ket{b}\bra{b}Y_{1,-1}\ket{a},
\label{angular_integral}
\end{align}
which results in
\begin{align}
I &= \frac{(-1)^{2(2n-3)}(2n-1)(2n-3)(2n-2)!(2n-4)!}{2^{2(2n-3)}\left((n-1)!(n-2)!\right)^2} \\
&\times
\left(\frac{1}{4\pi}\right)^2 \left(-\frac{3(2\pi)^2}{8\pi \sqrt{6}}\right)\left(\frac{\sqrt{\pi}\Gamma(n)}{\Gamma(n+\frac{1}{2})}\right)^2,
\nonumber
\end{align}
where $\Gamma(n)$ is the Gamma function. Using $Y_{2,0}^{*}(\theta_R=0,\varphi) =\sqrt{\frac{5}{4\pi}} $ and the radial matrix element
\begin{align}
d_{n (n-1);n (n-2)} &= -\frac{3n}{2} \sqrt{(2n-1)},
\end{align}
we finally reach
\begin{align}
{\cal M} &= -\frac{8\pi}{R^3}\sqrt{\frac{2\pi}{15}}\sqrt{\frac{5}{4\pi}}\left(\frac{-3n}{2} \sqrt{(2n-1)} \right)^2 \\ &\times
\nonumber
 \frac{(-1)^{2(2n-3)}(2n-1)(2n-3)(2n-2)!(2n-4)!}{2^{2(2n-3)}\left((n-1)!(n-2)!\right)^2} \\ &\times
\nonumber
 \left(\frac{1}{4\pi}\right)^2  \left(-\frac{3(2\pi)^2}{8\pi \sqrt{6}}\right)\left(\frac{\sqrt{\pi}\Gamma(n)}{\Gamma(n+\frac{1}{2})}\right)^2.
\end{align}
See \cite{xia:circgate} for analytical results for the $\ket{aa}\leftrightarrow\ket{cc'}$ dipole-matrix elements, where $\ket{c'}=\ket{n-1,n-2,n-2}$.

\section{Classical Simulations of Rydberg Dimer}
\label{appendix_classical}

In the classical simulations, we adopted the Bohr-Sommerfeld atomic model to mimic the orbital behavior by using elliptical orbits for a classical point electron.
Initial positions and velocities are drawn from a random distribution that respects the target quantum numbers via energy and (angular momentum):
\begin{equation}
    E_n=-\frac{e^4m_e}{32\pi^2\epsilon_0^2\hbar^2}\frac{1}{n^2},
\end{equation}
\begin{equation}
    L_m=\hbar\sqrt{l(l+1)},
\end{equation}
where $m_e$ is the mass of the electron.

In the model, the electron follows an elliptic path and the semi-major ($A_n$) and semi-minor($B_{nl}$) axes are defined as:
\begin{equation}
    A_n = \frac{4\pi\epsilon_0\hbar^2}{m_e e^2}n^2, \;\;\;\;\;\;\;\;\;\; B_{nl}=\frac{l}{n}A_n.
\end{equation}

In the simulation, nuclei of the atoms are assumed to be motionless and the equation of motion for the electrons is

\begin{multline}
    \ddot{\mathbf{r}}_{ei}=-\frac{e^2}{4\pi\epsilon_0m_e}\Bigg( \frac{\mathbf{r}_{ei}-\mathbf{r}_{ni}}{|\mathbf{r}_{ei}-\mathbf{r}_{ni}|^3} \\+ \frac{\mathbf{r}_{ei}-\mathbf{r}_{n(i+1)}}{|\mathbf{r}_{ei}-\mathbf{r}_{n(i+1)}|^3} - \frac{\mathbf{r}_{ei}-\mathbf{r}_{e(i+1)}}{|\mathbf{r}_{ei}-\mathbf{r}_{e(i+1)}|^3}\Bigg),
\end{multline}
where the index $ni$ is the i$^{\text{th}}$ nucleus and the index $ei$ is the i$^{\text{th}}$ electron. The notation $(i+1)$ pertains here simply to the adjacent atom in a dimer.

The classical simulation is conducted by numerical evaluation of the equation of motion and averaging the results over random initial positions of the electron on the elliptic orbit. For this we vary in particular the relative orbital phase between the electrons, $\varphi_2-\varphi_1$, see \frefp{geometry_states}{b}.

The black dashed lines in \frefp{transport}{a} show finally the ensemble averaged angular momenta $L_k=|\overline{\bv{L}_k}|$, where $\bv{L}_k$ is the angular momentum of electron $k$ with respect to nucleus $k$. The model could be made more sophisticated by incorporating also the out of plane distribution of the Rydberg electron evident in \frefp{geometry_states}{b} or nuclear motion.

\section{Parameter constraints for Rydberg aggregates}
\label{appendix_constraints}

For the parameter space survey in \sref{param_regimes} we have utilised the following mathematical criteria to define when a
one-dimensional circular Rydberg atom chain can constitute a useful flexible Rydberg aggregate. We are following the approach of \cite{wuester:review}.

{\it Validity of the essential state model}: We have seen in \fref{spaghetti} that the essential state models based on $\ket{a}$, $\ket{b}$ or $\ket{a}$, $\ket{c}$ breaks down once adjacent n-manifolds begin to mix. We have taken the corresponding distance $\sub{d}{min}$ as the one where $C_3^{(ac)}(n)/\sub{d}{min}^3=1/(2n^2) - 1/(2(n+1)^2)$ (atomic units).\\
{\it Static aggregates:} From \bref{Vdd} we can infer a transfer time (Rabi oscillation period) $\sub{T}{hop}=\pi d^3/C_3$ for an excitation to migrate from a given atom to the neighboring one, if the inter-atomic spacing is $d$. We have calculated the corresponding time for $\sub{N}{hops}=100$ such transfers, given by $\sub{T}{trans} =\sub{N}{hops} \sub{T}{hop}$, imagining migration along an entire aggregate. We finally require $\sub{T}{trans}$ to be short compared to the system lifetime, which is determined for circular states based on \eref{taulive}.\\
{\it Perturbing acceleration:} The characteristic time for atom acceleration is $\sub{T}{acc}=\sqrt{\frac{d^5 \sub{m}{Rb}}{6 C_3^{(ac)}}}$ \cite{wuester:review}, with mass of the atoms $ \sub{m}{Rb}$ and their initial separation $d$. We then color the parameter space red in \fref{circagg_regimes}, where atoms would inadvertently be set into motion due to $4\sub{T}{acc}<\sub{T}{trans}$ .\\
{\it Flexible aggregates:} For flexible aggregates, we assume an equidistant chain with spacing $d$, but the existence of a dislocation on the first two atoms with spacing of only $\sub{d}{ini}=a=d/2$ to initiate directed motion, similar to \sref{tully}. Hence, $\sub{d}{ini}>\sub{d}{min}$ must be fullfilled, a tighter constraint than $d>\sub{d}{min}$. We can then assess as in \cite{wuester:review} whether an excitation transporting pulse can traverse the chain within the system lifetime.

\acknowledgments
We gratefully acknowledge fruitful discussions with Mehmet Oktel and Michel Brune.


\begin{thebibliography}{52}
\bibitem[{foo()}]{footnote:space_reduction}
\bibinfo{note}{We collect all states that are coupled to the initial state via
  at most $c=3$ coupling matrix elements. Of these we remove all states that
  are detuned by more than $\sub{E}{cut}=2000$ MHz from the initial state.
  Finally we verified that results do not change when these numerical
  constraints were loosened.}
\end{thebibliography}
\end{document}